\documentclass[aps, prd, preprint,nobibnotes,nofootinbib]{revtex4-1}
\usepackage[utf8]{inputenc}
\usepackage{amsmath, amssymb, graphicx, mathrsfs, bbm, color, hyperref}

\newcommand{\beq}{\begin{equation}}
\newcommand{\eeq}{\end{equation}}

\newcommand{\eq}[1]{eq. (\ref{#1})}
\newcommand{\wt}[1]{\widetilde{#1}}

\newcommand{\tr}{\text{tr}}

\newcommand{\intg}{\mathbb{Z}}
\newcommand{\real}{\mathbb{R}}

\newcommand{\U}{\mathrm{U}}

\newcommand{\SO}{\mathrm{SO}}

\newcommand{\Spin}{\mathrm{Spin}}

\newcommand{\hlt}{\mathcal{H}}

\newcommand{\ve}{\varepsilon}

\newcommand{\vA}{\mathbf{A}}

\newcommand{\vp}{\mathbf{p}}

\newcommand{\vx}{\mathbf{x}}

\newcommand{\bpi}{\boldsymbol{\pi}}
\newcommand{\bps}{\boldsymbol{\psi}}

\newcommand{\bxi}{{\boldsymbol{\xi}}}
\newcommand{\bze}{{\boldsymbol{\zeta}}}

\newcommand{\bconn}{\mathfrak{a}}
\newcommand{\bcurv}{\mathfrak{F}}
\newcommand{\sform}{\rho}
\newcommand{\mcform}{\mathfrak{w}}
\newcommand{\vol}{\mathbb{V}}
\newcommand{\svol}{\Omega_H}

\newcommand{\emt}{\mathcal{T}}
\newcommand{\ccur}{J}
\newcommand{\ecur}{S}
\newcommand{\accur}{\mathcal{J}}
\newcommand{\aecur}{\mathcal{S}}
\newcommand{\anom}{\mathcal{A}}
\newcommand{\gibbs}{\mathcal{G}}

\newcommand{\bu}{\bar{u}}
\newcommand{\bq}{\bar{q}}
\newcommand{\bcc}{\bar{\mathcal{J}}}
\newcommand{\bec}{\bar{\mathcal{S}}}

\newcommand{\bg}{\bar{\mathcal{G}}}

\newcommand{\vel}{c}
\newcommand{\spin}{\mathfrak{S}}

\newcommand{\reg}{\mathscr{P}}
\newcommand{\phsp}{\mathcal{M}}
\newcommand{\orbit}{\mathcal{O}_\Lambda}

\begin{document}
\title{\bf Chiral kinetic theory and anomalous hydrodynamics in even spacetime dimensions}

\author{Vatsal Dwivedi}
\author{Michael Stone}
\affiliation{Department of Physics and Institute for Condensed Matter Theory, \\ University of Illinois at Urbana-Champaign, IL 61801, USA \\ \ \\}

\begin{abstract}
We study the hydrodynamics of a gas of noninteracting Weyl fermions coupled to the electromagnetic field in $(2N+1)+1$ spacetime dimensions using the chiral kinetic theory, which encodes the gauge anomaly in the Chern character of the nonabelian Berry connection over the Fermi surface. We derive the anomalous contributions to the relativistic hydrodynamic currents in equilibrium and at a finite temperature, which agree with and provides an approach complementary to the results derived previously using thermodynamic constraints.
\end{abstract}

\maketitle

% =========================== SECTION 1 ================================================================================================

\section{Introduction}
The equations of relativistic hydrodynamics, proposed in the 1940s\cite{eckart_rel_fluids, landau_fluids}, have been a versatile tool for the study of a wide variety of classical fluids. They have also been used to describe the long wavelength behavior of quantum field theories(QFTs), especially the usually intractable strongly interacting cases. A fluid description is useful as it can often be systematically constructed as a derivative expansion, knowing only the symmetries (and corresponding conservation laws) of the underlying QFT\cite{fluid-gravity}. 

Anomalies, the breakdown of a conservation law of a classical field theory when the theory is quantized\cite{bertlmann_anom}, are one of the most subtle and interesting aspects of quantum field theory. Their importance lies in their topological nature, as they correspond to the density of the topological index of a Dirac operator\cite{nakahara}. In principle, a theory with an anomalous dynamical gauge field is inconsistent, as the anomaly implies a breakdown of the gauge invariance; however, anomalies in a non-dynamical background gauge field can offer insightful diagnostics into the theory. 

The anomalies can have macroscopic consequences in the hydrodynamic regime of the QFT\cite{son-surowka,neiman-oz}, as seen in the investigation of the so-called ``anomalous fluids''. In Ref \onlinecite{son-surowka}, Son and Sur\'owka show that in $3+1$ spacetime dimensions, the presence of an anomalous conservation law for a $U(1)$ current necessitates adding terms to the constitutive relation at the first order in derivative expansion, which can be constrained using the second law of thermodynamics. Subsequently, Loganayagam\cite{loga_anom_transport} derived the general solutions to the second law constraint in arbitrary even spacetime dimensions. 

More recently, Loganayagam and Sur\'owka\cite{loga_weyl_gas} have conjectured a very powerful result for the hydrodynamic description of Weyl fermions in even spacetime dimensions. They argue that the anomalous contributions to the hydrodynamic equations can be derived from a ``Gibbs free energy current'' $\gibbs$. Furthermore, in $d$ spacetime dimensions, $\gibbs$ can be obtained by the replacements $F \to \mu$ and $\tr \left\{ R^{2n} \right\} \to 2 (2 \pi T)^{2n} \; \forall \, n \in \intg^+$ in the anomaly polynomial in $d+2$ dimensions, where $\mu$ is the chemical potential and $T$ the temperature, and $F$ and $R$ denote the Maxwell and Riemann curvatures, respectively. These replacement rules have been further studied\cite{loga_holo} in a holographic setting using the tools of fluid-gravity duality\cite{fluid-gravity}. 

An alternative perspective on anomalous QFTs is provided by the semiclassical approach, where one studies the dynamics of wavepackets treated as classical particles. The only quantum aspects are the coupling to the Berry connection and the $\hbar$ occurring in the phase space volume. In a Hamiltonian picture, the coupling to the Berry curvature leads to a nontrivial (``anomalous'')  symplectic form on the phase space, so that the position and momentum coordinates are no more conjugate\cite{son-yamamoto,horvathy_symp}. This view has proved particularly useful in condensed matter physics\cite{niu-elec,niu-bphase}, for instance, in the study of transport in Weyl semimetals\cite{son-spivak_wsm}. 

In Ref \onlinecite{misha_ckt}, Stephanov and Yin showed that a kinetic theory based on a semiclassical description of charged noninteracting Weyl fermions in $3+1$ spacetime dimensions  reproduces the Adler-Bell-Jackiw anomaly\cite{bertlmann_anom} correctly. Subsequently, their computation was generalized to nonabelian gauge anomalies\cite{vd-ms_nabel} in arbitrary even spacetime dimensions\cite{vd-ms_arbt} by constructing an anomalous symplectic form on an extended phase space, where the anomaly signals a breakdown of the Liouville's theorem. The formalism has also been used to describe the transport processes associated with gauge anomalies, for instance, the chiral magnetic effect(CME) and chiral vortical effect(CVE)\cite{misha_ckt, son-spivak_wsm}. 

In this paper, we study the hydrodynamics of a gas of charged noninteracting Weyl fermions in arbitrary even spacetime dimensions. Starting from a semiclassical microscopic description and assuming the system to be in thermodynamic equilibrium with the comoving frame, we derive the anomalous contributions to the hydrodynamic currents, which depend on the electromagnetic field and vorticity of the fluid. At a finite temperature, we include both positive and negative energy sectors to get a closed form expressions for the currents, which are identical to those obtained in Ref \onlinecite{loga_weyl_gas} using thermodynamic constraints. 

The rest of this paper is organized as follows: in Sec \ref{sec:ckt}, we review the anomalous symplectic form and the extended phase space proposed in Ref. \onlinecite{vd-ms_arbt}. In Sec \ref{sec:fluid}, we review the basics of relativistic hydrodynamics, including the differential form notation proposed in Ref. \onlinecite{loga_anom_transport}. In Sec \ref{sec:cur}, we set up the formalism to derive expressions for macroscopic currents using the anomalous symplectic form, using which  we derive the anomalous hydrodynamic currents in Sec \ref{sec:deriv}. Finally, we discuss our conclusions in Sec \ref{sec:disc}. In the Appendices, we review the Fermi-Dirac distribution and associated quantities in Appendix \ref{app:fermi}, derive the symplectic form in a noninertial reference frame in Appendix \ref{app:frame} and show that the comoving frame used in our calculation satisfies the no-drag condition\cite{misha_no_drag} in Appendix \ref{app:no_drag}

We follow the general relativity convention for the Minkowski metric, where $\eta^{\mu\nu} = \mathrm{diag}\{-1,1,\dots 1\}$ on $\real^{2N+1,1}$. The Greek indices ($\mu,\nu$) run over all the spacetime coordinates and the Latin indices from the middle of the alphabet $(i,j,k)$ run over only the space coordinates, with Einstein summation for repeated indices. We set $\hbar = c = 1$.

% =========================== SECTION 2 ================================================================================================

\section{Semiclassical description of Weyl fermions}    \label{sec:ckt}
We briefly review the semiclassical description of Weyl fermions\cite{misha_ckt, vd-ms_nabel, vd-ms_arbt} and the symplectic formulation thereof. 

\subsection{Extended phase space}      \label{sec:ckt_ext}
Consider a classical system on a $2M$-dimensional phase space $\phsp$ with a set of coordinates $\bze = \left( \zeta^1, \dots \zeta^{2M} \right)$. A generic action functional on this phase space is given by
\beq
S[\bze] = \int dt \left( \eta_i \left( \bze,t \right) \dot\zeta^i - \hlt(\bze, t) \right). 
\eeq 
The system is nonautonomous as $\eta$ is time-dependent, so that the standard symplectic formalism\cite{abraham-marsden} cannot be used. Instead, we extend the phase space to $\phsp_H = \phsp \times \real$, with the time coordinate $t \in \real$. The action can be written as a line integral of the so-called Liouville 1-form $\eta_H$ along the trajectory:
\beq 
S[\bze] = \int \eta_H, \quad \eta_H = \eta_i \left( \bze,t \right) d \zeta^i - \hlt(\bze, t) dt. 
\eeq 
The equation of motion can be written elegantly in a coordinate independent fashion in terms of a generalized symplectic form $\sform_H = d\eta_H$ as 
\beq 
i_V \sform_H = 0, \quad\quad V = \frac{\partial}{\partial t} + \dot \zeta^i \frac{\partial}{\partial \zeta^i}. 
\eeq
In other words, the ``symplectic form'' is now degenerate with corank 1, and the null direction gives us the trajectory in the extended phase space. This setup has also been studied under the name of \emph{contact structure}, with the above equation due to \'Elie Cartan (For instance, see Ref \onlinecite{abraham-marsden}, Theorem 5.1.13). 

Explicitly, 
\beq 
\sform_H = \frac{1}{2} \left( \frac{\partial \eta_j}{\partial \zeta^i} - \frac{\partial \eta_i }{\partial \zeta^j} \right) d\zeta^i \wedge d\zeta^j 
- \left( \frac{\partial \eta_i }{\partial t} + \frac{\partial \hlt}{ \partial \zeta^i} \right) d\zeta^i \wedge dt. 
\eeq
so that the equation of motion, $i_V \sform_H = 0$, demands that
\begin{align*}
 \dot\zeta^i \left( \frac{\partial \eta_j}{\partial \zeta^i} - \frac{\partial \eta_i }{\partial \zeta^j} \right) + \left( \frac{\partial \eta_j }{\partial t} + \frac{\partial \hlt}{ \partial \zeta^j} \right) = & \; 0 \\ 
 \dot\zeta^i \left( \frac{\partial \eta_i }{\partial t} + \frac{\partial \hlt}{ \partial \zeta^i} \right) = & \; 0 
\end{align*}
The former is precisely the Euler-Lagrange equation $\delta S[\bze] = 0$, while the latter can be obtained from the former by multiplying with $\dot\zeta^j$. 

The extended phase space is also equipped with a symplectic volume form
\beq
\svol = \frac{1}{M!} \sform_H^{M} dt = \sqrt{\sform} \left( \bigwedge_{i=1}^{2M} d\zeta^i \right) \wedge dt.    \label{eq:svol}
\eeq
Liouville's theorem, the statement that the phase space volume is conserved under Hamiltonian flows, is simply $\mathcal{L}_V \svol = 0$, where $\mathcal{L}_V = i_V d + d i_V$ is the Lie derivative.

\subsection{Anomalous symplectic form}
In Ref \onlinecite{misha_ckt}, Stephanov and Yin showed that in $3 + 1$ spacetime dimensions, positive chirality Weyl fermions with charge $q$ coupled to a background electromagnetic field can be described by the classical action 
\beq 
S[\vx,\vp] = \int dt \left( \vp \cdot \dot\vx - \ve - q\phi + q \vA \cdot \dot\vx - \bconn \cdot \dot\vp \right),
\eeq 
where $\ve = \vel|\vp|, \, c = \pm 1$ is the energy of the particle. The electromagnetic field is minimally coupled, with $\phi$ and $\vA$ being the electromagnetic scalar and vector potential, respectively. The quantum effects at $O(\hbar)$ are encoded in the Berry connection, $\bconn$, which can be thought of as a $\U(1)$ gauge field on the momentum space.

We use the extended phase space formalism described Sec \ref{sec:ckt_ext} to define the Liouville 1-form\footnote{For Weyl fermions, $\bconn$ corresponds to a monopole field, with the monopole located at the band-touching point. Thus, $\eta_H$ is not globally well-defined.}
\beq 
  \eta_H = p_i dx^i - \vel |\vp| dt + q A - \bconn,
\eeq 
where we have defined the 1-forms $A = A_\mu dx^\mu = - \phi dt + \vA_i dx^i$ and $\bconn = \bconn_i dp^i$. The corresponding symplectic form is 
\beq 
  \sform_H \equiv d \eta_H =  d p_i \wedge dx^i - \vel \, d |\vp| \wedge dt + q F - \bcurv,
\eeq 
where $F = dA = \frac{1}{2} F_{\mu\nu} dx^\mu \wedge dx^\nu$ and $\bcurv = d \bconn = \frac{1}{2} \bcurv_{ij} dp^i \wedge dp^j$. 

In $3+1$ dimensions, the Berry connection is abelian, since the Weyl fermions have only two components so that the positive energy sector is described by a single component. However, in $2N + 2$ spacetime dimensions, the positive energy sector is $2^{N+1}$-fold degenerate, so that the Berry connection becomes nonabelian with the gauge group $\Spin(2^{N+1})$. In Ref. \onlinecite{vd-ms_arbt}, we include this nonabelian Berry connection in the classical description by ``dequantizing'' it \emph{\`a la} Wong\cite{wong_dequant}. Given a representation of a compact gauge group $G$ with the highest weight vector $\Lambda$, one chooses an element $\alpha_\Lambda$ in the Cartan subalgebra of the Lie algebra, $\mathfrak{g}$. The classical description then involves enlarging the phase space to include the adjoint orbit\cite{kirillov_orbit} of $\alpha_\Lambda$, denoted by $\orbit$. 

Explicitly, we can define coordinates on $\orbit$ as $\spin = \sigma \alpha_\Lambda \sigma^{-1}, \; \sigma \in G$. Clearly, $\spin$ is invariant under $\sigma \to \sigma \cdot \eta \; \forall \eta \in H$, where  $H \subset G$ is the subgroup generated by the elements of Lie algebra that commute with $\alpha_\Lambda$, so that the orbit can be identified with the quotient $G/H$. Choose a basis  $\{\lambda_a\}$ of $\mathfrak{g}$, which satisfies the orthonormality condition $\tr_\Lambda \{ \lambda_a \lambda_b \} = \delta_{ab}$. Then, $\spin \in \mathfrak{g}$, being simply an adjoint action on $\alpha_\Lambda \in \mathfrak{g}$, can be written as $\spin = \spin^a \lambda_a$. Similarly, $\bconn = \bconn^a \lambda_a$ and $\bcurv = \bcurv^a \lambda_a$. For the ``dequantization'', the matrix-valued gauge connection and curvature are then replaced by
\begin{align}
\bconn & \; \mapsto \wt{\bconn} \equiv \tr \{\spin \bconn\} = \spin^a \bconn_a \nonumber \\
\bcurv & \; \mapsto \wt{\bcurv} \equiv \tr \{\spin \bcurv \} = \spin^a \bcurv_a.  
\end{align}
To make $\spin$, dynamical, we add the corresponding right Maurer-Cartan form $\mcform_R = d\sigma \sigma^{-1}$ to the Liouville 1-form. Thus, in $2N+2$ spacetime dimensions, our Liouville 1-form on the extended phase space $\phsp_H = \real^{4N+3} \times \orbit$ becomes 
\beq 
\eta_H =  p_i dx^i - \vel \, |\vp| dt + q A - \tr\{ \spin \left( \bconn + i \mcform_R \right) \}.
\eeq
The corresponding symplectic form is 
\beq 
  \sform_H \equiv d \eta_H =  d p_i \wedge dx^i - \vel \, d |\vp| \wedge dt + q F - \wt{\bcurv} - i \, \tr\left\{ \spin \left( \mcform_R - i \bconn \right)^2 \right\} .
\eeq

% =========================== SECTION 3 ================================================================================================

\section{Anomalous fluids}    \label{sec:fluid}
The dynamics of an anomalous fluid with a $U(1)$ anomaly is described by 
\beq 
\partial_\mu \emt^{\mu\nu} = F^{\nu\lambda} \ccur_\lambda, \quad \partial_\mu J^\mu = \anom, 
\eeq
where $\emt^{\mu\nu}$ is the energy-momentum tensor (``energy current'') of the fluid, $\ccur^\nu$ is the charge current, $F_{\mu\nu}$ is the Maxwell gauge field corresponding to the gauge connection $A_\mu(x)$ and $\anom(F)$ is the anomaly polynomial. One also defines an entropy current $\ecur^\mu$, which must satisfy the second law of thermodynamics, $\partial_\mu S^\mu \geq 0$.

In order to obtain a closed system of equations, the hydrodynamic currents need to be expressed in terms of the thermodynamic fields, viz, the velocity $u^\mu(x)$ (satisfying $u_\mu u^\mu = -1$), the temperature $T(x)$, the chemical potential $\mu(x)$ and the gauge connection $A_\mu(x)$. These are the so-called \emph{constitutive relations}, which can be constructed systematically in a derivative expansion, where each spacetime derivative of the thermodynamic fields counts as dimension 1 for the bookkeeping\cite{fluid-gravity}.

For anomalous and dissipationless fluids, the most general\footnote{In principle, there can also be tensor corrections to $\emt$ due to the anomaly. They are usually ignored in thermodynamic calculations, as they cannot be constrained by the second law of thermodynamics\cite{loga_anom_transport}.}  
constitutive relations can be written as\cite{son-surowka, loga_anom_transport, loga_weyl_gas}
\begin{align}
 \emt^{\mu\nu} = & \, (\ve + p) u^\mu u^\nu + p \eta^{\mu\nu} + (q^\mu u^\nu + u^\mu q^\nu), \nonumber \\ 
 \ccur^\mu =  & \, n u^\mu + \accur^\mu, \nonumber \\ 
 \ecur^\mu =  & \, s u^\mu + \aecur^\mu,    \label{eq:const}
\end{align}
where $q, \accur$ and $\aecur$ contain one or more spacetime derivatives of $u_\mu$ or $A_\mu$. We also set 
\beq  
 u_\mu q^\mu  = u_\mu \accur^\mu  = u_\mu \aecur^\mu = 0,   \label{eq:trans_cond}
\eeq
so that in the frame where $u^\mu = (1, 0, \dots 0)$, the components $\emt^{00} \equiv \ve$ and $J^0 \equiv n$ represent the actual energy density and charge density, respectively.  

At the first order in a derivative expansion, we can define the ``curvatures''
\beq 
F_{\mu\nu} = \partial_\mu A_\nu - \partial_\nu A_\mu, \quad \Omega_{\mu\nu} = \partial_\mu u_\nu - \partial_\nu u_\mu. 
\eeq 
In a frame specified by $u$, the electric field is defined as $E_\mu = F_{\mu\nu} u^\nu$. Since the acceleration is defined as $a_\mu = u^\nu \partial_\nu u_\mu$, we can use $u^\nu \partial_\mu u_\nu = \frac{1}{2} \partial_\mu\left( u_\nu u^\nu \right) = 0$ to derive an analogous expression
\beq 
a_\mu = u^\nu \left( \partial_\nu u_\mu - \partial_\mu u_\nu \right) = - \Omega_{\mu\nu} u^\nu. 
\eeq 
It is convenient to rephrase the above expressions in the language of differential forms. We define the 1-forms $u = u_\mu dx^\mu$ and $A = A_\mu dx^\mu$, and decompose their exterior derivatives into the ``magnetic'' and ``electric'' components\cite{loga_anom_transport} as
\begin{align}
 F = & \; dA = B + u \wedge E, \nonumber \\ 
 \Omega = & \; du \, = \omega - u \wedge a, 
\end{align}
where $B$ and $\omega$ are 2-forms. Clearly, 
\beq 
u \wedge F = u \wedge B, \quad\quad u\wedge \Omega = u \wedge \omega.    \label{eq:u_wedge_F}
\eeq
Note that in order to expose the mathematical symmetry between these decompositions, we have followed the traditional fluid mechanics convention\cite{landau_fluids} in defining the \emph{vorticity} $\omega$, so that $\vec{\omega} = \nabla \times \vec{u}$ in 3+1 dimensions. This is in contrast to the angular velocity, $\vec{\omega}_{A} = \frac{1}{2} \nabla \times \vec{u} = \frac{1}{2} \omega$, which is sometimes referred to as the ``vorticity'' in relativistic hydrodynamics (for instance, Refs \onlinecite{fluid-gravity, loga_anom_transport}). 

The decompositions are particularly transparent in a frame where $u^\mu = \left( 1, 0, \dots 0 \right)$, so that $u \equiv u_\mu dx^\mu =  -dt$. Then, $E_0 = 0, \; E_i = F_{i 0}$ and $a_0 = 0, \; a_i = -\Omega_{i0} = \partial_t u_i$, so that 
\begin{align*}
 F = & \; B - F_{i0} \, dt \wedge dx^i \implies B = \frac{1}{2} F_{ij} dx^i \wedge dx^j \\ 
 \Omega = & \; \omega - \Omega_{i0} \, dt \wedge dx^i \implies \omega = \frac{1}{2} \Omega_{ij} dx^i \wedge dx^j =  \partial_i u_j dx^i \wedge dx^j. 
\end{align*}

To express the conservation laws in the language of differential forms, we also need the Hodge dual\cite{stone-MM}. We follow Loganayagam\cite{loga_anom_transport} in denoting the Hodge dual by an overbar as well as the usual $\star$. For instance, in $3 + 1$ dimensions, 
\beq 
\bu = u_\mu (\star dx^\mu) = \frac{1}{3!} u_\mu \epsilon^\mu_{\phantom{\mu}\nu\rho\lambda} dx^\nu dx^\lambda dx^\rho = \left( \frac{1}{3!} \epsilon_{\mu\nu\rho\lambda} u^\mu \right) dx^\nu dx^\lambda dx^\rho.
\eeq
Given 1-forms $u$ and $v$, the inner product and gradient can be written as 
\beq 
\star\, (v_\mu u^\mu) = v \wedge \bu, \quad \star\, (\partial_\mu u^\mu) = d\bu, 
\eeq 
where $\star 1 =  \vol$, the Euclidean volume form on $\real^{2N+1,1}$. This is the dictionary to go between the differential forms and vectors on $\real^{2N+1, 1}$. 

Finally, we  define the 1-forms corresponding to the anomalous currents as $q = q_\mu dx^\mu$, etc and their Hodge duals as $\bq = q_\mu \left( \star dx^\mu \right)$, which are $(2N+1)$-forms. Schematically, we can construct such a form by wedging the 1-form $u$ to $N-1$ copies of $F$ or $\Omega$, i.e, 
\beq
\bq \sim \sum_{k=1}^{N-1} \xi_{q, k} \, u \wedge \underbrace{F \wedge \dots \wedge F}_{k} \wedge \underbrace{\Omega \wedge \dots \wedge \Omega}_{N-1-k},   \label{eq:tr_coeff}
\eeq 
with similar expressions for $\bcc$ and $\bec$. The task of hydrodynamics is then to constrain the transport coefficients ($\xi$'s here) using general principles such as those of thermodynamics. 

The authors of Ref. \onlinecite{loga_anom_transport} also define a grand potential\footnote{In Ref \onlinecite{loga_weyl_gas}, $\gibbs$ is referred to as the Gibbs free energy current. However, as the Gibbs free energy (per unit volume) is defined as $G = \ve + p - Ts = \mu n$, the ``Gibbs free energy current'' would be defined as $\mu J$. } 
current $\bg$, which acts as a generating function for $\bq$, $\bcc$ and $\bec$:
\beq 
\bg = \bq - \mu \bcc - T \bec; \quad\quad \bcc = -\frac{\partial \bg}{\partial \mu}, \quad \bec = -\frac{\partial \bg}{\partial T}. 
\eeq 
We shall derive this current by an explicit semiclassical calculation.

% =========================== SECTION 4 ================================================================================================

\section{Symplectic form and currents}     \label{sec:cur}
In Sec \ref{sec:ckt}, we reviewed the semiclassical Hamiltonian description of Weyl fermions in an inertial reference frame. However, for hydrodynamics, it is more natural to consider the co-moving frame, defined by the given velocity field $u^\mu(x)$. As the frame may in general possesses a nonzero acceleration as well as vorticity $(\Omega = du \neq 0)$, we need a way to include the inertial forces in our formalism.  

In Appendix \ref{app:frame}, we derive the generalized symplectic form in a noninertial reference frame, and show that for massless particles, it is reasonable to include the inertial forces in the symplectic form as $\sform_H \to \sform_H + \ve \Omega$ (at the linear order in $\Omega$). This is reminiscent of the minimal coupling to the electromagnetic field, with $\ve = \vel|\vp|$ serving as the ``charge''. Thus, the semiclassical dynamics of Weyl fermions in the co-moving frame on $\real^{2N+1,1}$ is described by the generalized symplectic form 
\beq 
  \sform_H \equiv d \eta_H =  d p_i \wedge dx^i - \vel\, d |\vp| \wedge dt + q F + \vel\, |\vp|\Omega - \wt{\bcurv} - i \, \tr\left\{ \spin \left( \mcform_R - i \bconn \right)^2 \right\}, \label{eq:sform_final}
\eeq 
where we have locally set $u^\mu = (1,0, \dots 0)$ by suitable Lorentz transforms, so that $-dt = u$. The $(2M+1)$-dimensional extended phase space is $\phsp_H = \real^{2n} \times \real \times \orbit, \, n = 2N+1$, where $M = n + \frac{1}{2} \text{dim}\left( \orbit \right) \equiv n + m_\Lambda$.

Consider now the space density of a energy-dependent physical quantity, $Q(\ve,x)$. The corresponding current is defined as 
\beq 
J_c^i = \int_{\reg} \, \frac{d^np}{(2\pi)^n}  \, d\mu_\Lambda \, \sqrt{\rho}\, \dot x^i \, Q(\ve,x),
\eeq 
where $\reg = \real^n \times \orbit$ and $d\mu_\Lambda$ is an invariant measure on $\orbit$. In principle, one next needs to solve the equation of motion $(i_V \sform_H = 0)$ for $\sqrt{\sform} \, \dot x^i$, which can then be integrated over the momentum space and the co-adjoint orbit. In 3+1 dimensions, this is straightforward\cite{misha_ckt}, and one gets
\beq 
\sqrt{\sform} \, \dot \vx = \vel\,\hat{\vp} + \mathfrak{b} \times \mathbf{E} + \left( \hat{\vp} \cdot \mathfrak{b} \right) \left( q \mathbf{B} + \vel |\vp| \boldsymbol{\omega} \right), \quad\quad\quad \mathfrak{b}^i = \frac{1}{2} \epsilon^{ijk} \bcurv_{jk}.
\eeq 
However, the task is much more complicated in spacetime dimensions greater than 4. Thus, we follow an alternative approach using the symplectic formulation of classical mechanics, which lets us compute such currents without computing $\sqrt{\rho} \, \dot x^i$ explicitly. 

Define the current 1-form $J_c = J_{c,i} dx^i$, whose Hodge dual can be explicitly written as 
\beq 
 \bar{J}_c = \frac{1}{(2\pi)^n}  \int_\reg Q(\ve,x) \sqrt{\sform} \left( \sum_{i=1}^n (-1)^i dx^1 \wedge \dots dx^{i-1} \wedge \dot x^i dx^{i+1} \wedge \dots dx^{n} \right) \frac{d^np}{(2\pi)^n} \wedge dt \wedge d\mu_\Lambda,
\eeq 
The differential form in the parenthesis is simply 
\[
i_V \left( d^{n} x  \right) = i_V  \left( \bigwedge_{i=1}^n dx^i \right) = \sum_{i=1}^n (-1)^i dx^1 \wedge \dots dx^{i-1} \wedge \dot x^i dx^{i+1} \wedge \dots dx^{n},
\]
so that the measure on the right hand side can readily be obtained as an antiderivation of the symplectic volume form, with only one term at $O(d^{n-1}x)$:
\begin{align}
 i_V \svol = & \; i_V \left( \sqrt{\sform} \, d^nx \wedge d^np \wedge dt \wedge d\mu_\Lambda \right) \nonumber \\ 
 = & \; \sqrt{\rho} \left[ i_V \left( d^{n}x \right) \wedge d^n p \wedge dt \wedge d\mu_\Lambda + \text{terms involving }d^nx \right].
\end{align}
Using \eq{eq:svol} and our explicit expression for the symplectic form in \eq{eq:sform_final}, we can explicitly write the symplectic volume form as
\beq 
\Omega_H = \frac{1}{M!} \sform_H^M dt =  \frac{1}{n!} \sform_0^n  dt \, d\mu_\Lambda, \quad
d\mu_\Lambda = \frac{1}{m_\Lambda!} \left[ - i \tr \left\{ \spin \mcform_R^2 \right\} \right]^{m_\Lambda}
\eeq 
where we have defined 
\beq
 \sform_0 = d p_i \wedge dx^i - \vel\, d |\vp| \wedge dt + qF + \vel |\vp| \Omega - \wt{\bcurv},
\eeq
which contains only the spacetime and momentum differentials. Thus, using $i_V dt = 1$ and the equation of motion $i_V \sform_H = 0$, we get
\beq
 i_V \svol = \frac{1}{n!} i_V \left[ \sform_0^{n}  \, dt \, d\mu_\Lambda\right] = \frac{1}{n!} \left[ \sform_0^n d\mu_\Lambda + \sform_0^n \, dt (i_V d\mu_\Lambda ) \right].
\eeq
As we seek the terms at $O(d^{n-1}x)$, we simply need to read off the coefficient of $\left( i_V d^nx \right) d^np \, dt$ in $\frac{1}{n!} \rho_0^n$.  We obtain terms of the form
\[
  \left(  d p_i \wedge dx^i - \vel\, d |\vp| \wedge dt \right)^{2(N-k)+1} \left( qF + \vel |\vp| \Omega \right)^k \wedge \left( \wt{\bcurv} \right)^k; \quad 0 \leq k \leq N. 
\]
To compute $\bar{J}_c$, we need to integrate these terms over $\reg = \real^{2N+1} \times \orbit \cong \real^+ \times S^{2N} \times \orbit$, where $\real^+$ denotes the radial $|\vp|$ axis.
We now show that only the term with $k = N$ integrates to a nonzero value over $\reg$. 

For the Weyl Hamiltonian $\hlt = \vp \cdot \Gamma$, the Berry curvature is singular at the band touching point $\vp = 0$, which acts as a nonabelian monopole in the Berry curvature field\cite{vd-ms_arbt}. Mathematically, the states  correspond to a complex line bundle $\mathfrak{C}$ over the unit sphere in momentum space, which can be written as a direct sum of the subbundles corresponding to positive and negative energies. The positive energy subbundle carries a Chern number $\chi$, equal to the chirality of the node, and as $\mathfrak{C}$ is trivial, the negative energy subbundle carries a Chern number $-\chi$ (also see Ref. \onlinecite{vd-ms_arbt}, eq B9). As we are only considering a positive chirality Weyl node, we set $\chi = +1$. 

Thus, using the definition of Chern number\cite{nakahara} and $\wt{\bcurv} = \spin^a \bcurv_a$, we get
\begin{align}
\frac{(-1)^N}{N! (2\pi)^N} \int_{S^{2N} \times \orbit} \wt{\bcurv}^N \wedge d\mu_\Lambda 
= & \; \frac{(-1)^N}{N! (2\pi)^N} \int_{S^{2N}} \bcurv^{a_1} \dots  \bcurv^{a_N} \int_{\orbit} d\mu_\Lambda \spin_{a_1} \dots \spin_{a_N} \nonumber \\ 
= & \; \frac{(-1)^N}{N! (2\pi)^N} \int_{S^{2N}} \bcurv^{a_1} \dots  \bcurv^{a_N} \tr\{ \lambda_{a_1} \dots \lambda_{a_N} \} \nonumber \\ 
= & \; \frac{1}{N!} \int_{S^{2N}} \tr \left\{ \left( - \frac{\bcurv}{2\pi} \right)^N \right\} = \vel, \label{eq:chern}
\end{align}
where $\vel = \pm 1$ for the positive/negative energy subspace. In the first line, we have replaced the integral of $\spin$'s over $\orbit$, the classical phase space, with a trace of a product of generators over the quantum representation. This is only an approximation, which can be improved by integrating over $\mathcal{O}_{\Lambda + \mathfrak{W}}$ instead of $\orbit$, where we have shifted the highest weight vector $\Lambda$ by the Weyl vector $\mathfrak{W}$ (See Appendix C of Ref. \onlinecite{vd-ms_arbt} for details). Better still, we can ``requantize'' the co-adjoint orbit to reproduce the quantum traces, using the Borel-Weil-Bott construction (Ref \onlinecite{stone-MM}, Sec 16.2.3).

Next, we note that any terms with $0<k<N$ integrate to zero, since in order for $\tr \{\bcurv^k\} \wedge d^{2(N-k)}p$ to integrate to a nonzero value over $S^{2N}$, we need to integrate $\tr \{\bcurv^k\}$ over a nontrivial $2k$-cycle in $S^{2N}$. However, the only nontrivial cycles in $S^{2N}$ are in dimensions $2N$ and zero\cite{nakahara, stone-MM}, so that all such integrals with $0 < k < N$ evaluate to zero. 

Finally, for $k = 0$, considering the integral for $J^i$ and using $d|\vp| = \hat{p_j} dp^j$, the integral over the momentum space $\real^n, \; n = 2N+1$, is 
\[
\int_{\real^{n}} Q(\vel|\vp|,x) \hat p_j dp^j \left(\bigwedge_{\ell \neq i} dp^\ell \right) = \int_{\real^{n}} \hat p_i  Q(\vel|\vp|,x) \left(\bigwedge_{\ell=1}^n dp^\ell \right), 
\]
which vanishes, as the integrand is odd under $p_i \to -p_i$. Thus, $J_c$ would involve only the integral of
\[ 
\frac{1}{(N!)^2} (- \vel \, d|\vp| \wedge dt) \wedge (qF + \vel |\vp|\Omega)^N \wedge \left( -\wt{\bcurv} \right)^N 
\] 
over $\reg$. Explicitly, 
\beq 
 \bar{J}_c = -  \frac{\vel}{2\pi (N!)^2} \int_\reg Q(\vel|\vp|,x)  d|\vp| \wedge \left( - \frac{\wt{\bcurv}}{2\pi} \right)^N \wedge d\mu_\Lambda \wedge dt \wedge \left( \frac{qF + \vel |\vp|\Omega}{2\pi} \right)^N. 
\eeq 
Integrating over $S^{2N} \times \orbit$ using \eq{eq:chern}, we get 
\beq
 \bar{J}_c = \frac{\vel^2}{N!} (-dt) \wedge \int_0^\infty \frac{d|\vp|}{2\pi} Q(\ve,x)  \left( \frac{qF + \vel |\vp|\Omega}{2\pi} \right)^N .  
\eeq 
Finally, substituting $u = -dt$ and $\vel^2 = 1$ and using \eq{eq:u_wedge_F}, 
\beq
 \bar{J}_c =  \frac{1}{N!} \, u \wedge \int_0^\infty \frac{d|\vp|}{2\pi} Q(\vel|\vp|,x) \left( \frac{qB + \vel|\vp|\omega}{2\pi} \right)^N.       \label{eq:cur_final}
\eeq 
This is an explicit expression for the contribution of one energy sector ($c = \pm 1$) of a positive chirality Weyl node to the current $J$ in arbitrary even spacetime dimensions. In the next section, we derive the relevant $Q(\ve,x)$ for the grand potential current in relativistic hydrodynamics of anomalous fluids.

% =========================== SECTION 5 ================================================================================================

\section{Microscopic derivation of hydrodynamic currents}  \label{sec:deriv}
Consider a gas of Weyl particles with positive chirality in the phase space in equilibrium with a given frame of reference, so that the phase space distribution is simply the Fermi-Dirac distribution
\beq 
f(p, x) \equiv f(\ve) =  \frac{1}{1 + e^{\beta (\ve - \mu(x)) }}, \quad \beta = \frac{1}{T},
\eeq
Since the particles are fermions,  we can define the microscopic entropy density
\beq 
h(\ve) = -\sum_{\textrm{states}} p_i \log p_i = - f \log f - (1-f) \log(1-f). 
\eeq 
Given the trajectory of a single particle, $\dot x^i$, the number current and energy-momentum tensor are defined as\footnote{We have suppressed the Dirac delta functions localizing these quantities to the particle trajectory.}
\beq 
j^i = \dot x^i, \quad t^{i\nu} = \ve \dot{x}^i \dot{x}^\nu.
\eeq 
The anomalous hydrodynamic currents of \eq{eq:const} can then be defined simply by averaging over all particles. Following the definitions of Sec. \ref{sec:cur}, we fefine the anomalous currents for a given energy sector as\footnote{Strictly speaking, we should be subtracting off the ``normal'' contribution, i.e, replacing $\dot x^i \to \dot x^i -  \dot x_0^i$, where $x_0^i(t)$ solves the equation of motion with $ \bcurv = 0$. But as the system is in equilibrium in the given frame, the normal component vanishes. }
\begin{align}
 q^i = & \; \int_{\reg} \, \frac{d^np}{(2\pi)^n}  \, d\mu_\Lambda \, \sqrt{\rho}\, \dot x^i \, \ve \, f(\vel|\vp|),   \nonumber \\ 
 \accur^i = & \; \int_{\reg} \, \frac{d^np}{(2\pi)^n}  \, d\mu_\Lambda \, \sqrt{\rho}\, \dot x^i \, f(\vel|\vp|), \nonumber \\ 
 \aecur^i = & \; \int_{\reg} \, \frac{d^np}{(2\pi)^n}  \, d\mu_\Lambda \, \sqrt{\rho}\, \dot x^i \, h(\vel|\vp|),
\end{align}
where $q^i = \emt^{i0}$. As $u^\mu = \left( 1, 0, \dots 0 \right)$ and the anomalous contributions are transverse to $u$ (\eq{eq:trans_cond}), we also have $q^0 = \accur^0 = \aecur^0 = 0$. The corresponding grand potential current is given by 
\beq
 \gibbs_c^i = q^i - \mu \accur^i - T \aecur^i =  \int_{\reg} \, \frac{d^np}{(2\pi)^n}  \, d\mu_\Lambda \, \sqrt{\rho}\, \dot x^i g(\vel |\vp|),
\eeq 
where (using \eq{eq:fermi_g})
\beq 
(\ve - \mu) f(\ve) - T h (\ve)  = -\frac{1}{\beta} \ln\left(1 + e^{-\beta (\ve - \mu)} \right) \equiv g (\ve). 
\eeq
In order to derive physically meaningful expressions at a finite temperature, we must include both positive and negative energy sectors. Using \eq{eq:cur_final}, we define
\begin{align}
 \bg = & \; \bg_+ + \bg_- \nonumber \\ 
 = & \;  \frac{u}{N!} \wedge  \left[ \int_0^\infty \frac{d|\vp|}{2\pi} g(|\vp|) \left( \frac{qB + |\vp|\omega}{2\pi} \right)^N + \int_0^\infty \frac{d|\vp|}{2\pi}  g(-|\vp|) \left( \frac{qB - |\vp|\omega}{2\pi} \right)^N \right].
\end{align}
Substituting $|\vp| = \ve$ in the first integral and $|\vp| = -\ve$ in the second, we get 
\beq 
\bg =  \frac{u}{N!} \wedge  \int_{-\infty}^\infty \frac{d\ve}{2\pi} g(\ve) \left( \frac{qB + \ve\omega}{2\pi} \right)^N.   \label{eq:G_int}
\eeq 
This integral is clearly divergent, as $g(\ve) \sim (\ve - \mu)$ for $\ve \to -\infty$. This is expected, as we are integrating over an infinitely deep Dirac sea. In order to regularize this integral, we need to subtract off the zero temperature vacuum contribution, where we define the ``vacuum'' as the many-body state where all 1-particle states with $\ve < 0$ (i.e, below the Weyl node) are filled up. Since at $T = 0$,  $g(\ve) = (\ve - \mu) \Theta(\mu - \ve)$, where $\mu > 0$, define the regularized grand potential current as
\beq 
\bg^{reg} =  \frac{u}{N!} \wedge  \int_{-\infty}^\infty \frac{d\ve}{2\pi} \left[g(\ve) - (\ve - \mu) \Theta(-\ve) \right] \left( \frac{qB + \ve\omega}{2\pi} \right)^N.   \label{eq:G_reg_int}
\eeq 
Following Ref. \onlinecite{loga_weyl_gas}, we define a generating function $\bg^{reg}_\tau$ by multiplying the integral by $\tau^N$ and formally\footnote{i.e, we treat $\omega$ and $B$ as c-numbers instead of differential forms.} summing over $N$ to get
\beq 
\bg_\tau^{reg} = u \wedge e^{\frac{\tau q B}{2\pi}} \int_{-\infty}^\infty \frac{d\ve}{2\pi}  \left[g(\ve) - (\ve - \mu) \Theta(-\ve) \right] e^{\frac{\tau\ve\omega}{2\pi}}. 
\eeq 
We evaluate the integral explicitly (Appendix \ref{app:fermi}, \eq{eq:g_int}) to get 
\beq 
\bg_\tau^{reg} =  -u \wedge e^{\frac{\tau q B}{2\pi}} \frac{2\pi}{(\omega\tau)^2} \left[ \frac{\frac{\omega\tau}{2\beta}}{\sin\left( \frac{\omega\tau}{2\beta} \right)} e^{\frac{\mu\omega\tau}{2\pi}} - \left( 1 + \frac{\mu}{2\pi} \omega\tau \right) \right],
\eeq 
which is identical to the expression obtained\footnote{After replacing $\omega \to 2 \omega_{A}$, as they defines their vorticity as the angular velocity, $\omega_{A}$.} in Ref. \onlinecite{loga_weyl_gas}. In order to obtain the anomalous currents in $2N+2$ spacetime dimensions, $\left. \bg^{reg} \right|_{d=2N+2}$, we expand $\bg^{reg}_\tau$ in a power series in $\tau$ and pick out the coefficient of $\tau^N$. 

The generating function $\bg^{reg}_\tau$, as pointed out in Ref. \onlinecite{loga_weyl_gas}, is remarkable in its form, which is identical to that of the generating function for gauge and gravitational anomaly polynomials. An obvious but nontrivial consequence is that the transport coefficients, which can be derived using \eq{eq:tr_coeff}, are polynomials in $T$ and $\mu$ in all spacetime dimensions.

% =========================== SECTION 5 ================================================================================================

\section{Conclusion and Discussion}   \label{sec:disc}
We have derived the anomalous contributions to the relativistic hydrodynamic currents from a microscopic semiclassical description of Weyl fermions, which do agree with the corresponding expressions derived earlier using thermodynamic constraints\cite{loga_weyl_gas, misha_no_drag}. Starting from a semiclassical theory, our calculation exposes the role the (nonabelian) Berry curvature plays in the semiclassial dynamics of chiral fermions. The semiclassical formalism encodes the anomaly via the collisionless Boltzmann equation\cite{vd-ms_arbt}, $\mathcal{L}_V (f \Omega_H) \sim \anom$. Our approach, originally proposed to derive the anomalies from a semiclassical calculation, is complementary to the usual hydrodynamic approach, which takes the anomaly as given and explores its conseqences on the transport properties of the system. 

In order to compute the hydrodynamic currents at a finite temperature, we needed both positive and negative energy sectors of a \emph{single} positive chirality Weyl node. Even though the contribution of negative energy states is minuscule for $\mu \gg T$, it was required to obtain closed form expressions for the currents, where the transport coefficients turn out to be polynomials in $T$ and $\mu$. This is in contrast to the calculation in Ref \onlinecite{loga_weyl_gas}, where the contributions from the ``particle'' and ``antiparticle'' sectors were needed, which correspond to Weyl nodes of different chiralities in $4N$ spacetime dimensions.

A careful reader might have noticed that despite our claims to have derived the \emph{relativistic} currents, our expression for the symplectic form is not manifestly Lorentz invariant, as the definition of a Berry phase explicitly requires us to choose a foliation of the spacetime and to treat space and time on different footings. Indeed, a Lorentz invariant description has been attempted\cite{misha_ckt_lor, ms_lorentz}, leading to a nontrivial implementation of the Lorentz symmetry on our $x$ and $p$ variables as well as a magnetic moment correction to the energy. However, such corrections occur at $O(\hbar^2)$, so that our semiclassical symplectic form is accurate to $O(\hbar)$, and furthermore, the corrections do not affect the derivation of anomaly. Thus, we believe that the anomalous contributions obtained in this paper should be accurate to the lowest nontrivial order in $\hbar$ in a WKB-like expansion. 	

The semiclassical description used in this paper was proposed\cite{vd-ms_arbt} for Weyl fermions coupled to a nonabelian gauge field with a compact gauge group. Hence, the calculation presented in this paper readily generalizes to anomalous currents for hydrodynamics coupled to a nonabelian gauge field, provided the chemical potentials commute. However, as the nonabelian anomalies, being a breakdown of a covariant conservation law, cannot be interpreted as a spectral flow in any obvious way, the physical content of such a calculation is open to interpretation.

The similarity of the generating function for the grand potential current to the generating function for the gauge anomaly, and the corresponding replacement rules that it naturally suggests, is one of the many remarkable and mysterious results concerning anomalies. Whether the similarity points to a deeper connection (for instance, in holographic terms), or is a mere mathematical coincidence remains a very intriguing open question.

\acknowledgments 
This project was supported by the National Science Foundation under grant NSF DMR 13-06011.

% =========================== APPENDICES ================================================================================================

\appendix

\section{Fermi-Dirac distribution and integrals}     \label{app:fermi}
Consider a gas of fermions in the grand canonical ensemble. Owing to the Pauli exclusion principle, a given microstate can either be unoccupied or occupied by exactly one fermion. Thus, the 1-particle grand canonical partition function $z$ is 
\beq 
z = \sum_{\text{states}} e^{-\beta (\ve-\mu)} = 1 + e^{-\beta (\ve-\mu)},
\eeq 
where $\ve$ is the energy of the microstate, $\beta = T^{-1}$ is the inverse temperature and $\mu$ is the chemical potential. The corresponding grand potential $g$ is
\beq 
g = -\frac{1}{\beta} \ln z = -\frac{1}{\beta} \ln \left( 1 + e^{-\beta(\ve - \mu)} \right).
\eeq 
The grand potential is the generator for a variety of other relevant functions, such as the probability of occupation of a given state $f$ (Fermi-Dirac distribution) or the 1-particle entropy $h$:
\beq 
f = - \frac{\partial g}{\partial\mu} = \frac{1}{1 + e^{\beta(\ve - \mu)}}, \quad h = -\frac{\partial g}{\partial T}.
\eeq 
For fermions, we also have the highly nontrivial relation 
\beq 
 g = (\ve - \mu )f - T h,   \label{eq:fermi_g}
\eeq 
which can be obtained in a straightforward fashion using a definition of $h$ in terms of the occupation probability $f$:
\begin{align*}
 h 
 = & \; - \sum_{\text{states}} p_i \ln p_i = - f \ln f - (1-f) \ln(1-f) \\ 
 = & \;  \frac{1}{1 + e^{\beta (\ve - \mu)}} \ln\left( 1 + e^{\beta (\ve - \mu)} \right) + \frac{e^{\beta (\ve - \mu)}}{1 + e^{\beta (\ve - \mu)}} \ln\left( \frac{1 + e^{\beta (\ve - \mu)}}{e^{\beta (\ve - \mu)}} \right) \\ 
 = & \; \ln\left( 1 + e^{\beta (\ve - \mu)} \right) - \frac{e^{\beta (\ve - \mu)}}{1 + e^{\beta (\ve - \mu)}} \beta (\ve - \mu)  \\ 
 = & \; \ln\left(1 + e^{-\beta (\ve - \mu)} \right) + \beta (\ve - \mu) \left[1 - \frac{e^{\beta (\ve - \mu)}}{1 + e^{\beta (\ve - \mu)}} \right] \\ 
 = & \; \beta \left[ - g +  (\ve - \mu) f \right].
\end{align*}

\ \\ \noindent 
{\bf Integrals:} We seek to evaluate the integral
\[
 I(\sigma) = \int_{-\infty}^\infty \frac{d\ve}{2\pi} e^{\sigma\ve} \left[ g(\ve) -  (\ve - \mu) \Theta(-\ve) \right].
\]The Heaviside integral part is easy to evaluate: 
\beq 
\int_{-\infty}^0 \frac{d\ve}{2\pi} (\ve - \mu) e^{\sigma\ve} = \left( \frac{\partial}{\partial\sigma} -\mu \right) \int_{-\infty}^0 \frac{d\ve}{2\pi} e^{\sigma\ve} =  \left( \frac{\partial}{\partial\sigma} -\mu \right) \frac{1}{2\pi\sigma} = - \frac{1 + \mu\sigma}{2\pi\sigma^2}.    \label{eq:int2}
\eeq 
For the remaining integral, substitute $s = e^{\beta(\ve - \mu)} \implies ds = \beta s \, d\ve$ and integrate by parts to get
\begin{align}
 \int_{-\infty}^\infty \frac{d\ve}{2\pi} g(\ve) e^{\sigma\ve} 
 = & \; - \frac{1}{\beta} \int_0^\infty \frac{ds}{2\pi\beta\, s} \ln\left( 1 + \frac{1}{s} \right) \left( s\, e^{\beta\mu} \right)^{\frac{\sigma}{\beta}} \nonumber \\ 
 = & \; - \frac{e^{\mu\sigma}}{2\pi\beta^2} \int_0^\infty ds \, s^{\frac{\sigma}{\beta}-1} \ln\left( 1 + \frac{1}{s} \right) \nonumber \\ 
 = & \; - \frac{e^{\mu\sigma}}{2\pi\beta^2} \left[ \left. \frac{\beta}{\sigma}s^{\frac{\sigma}{\beta}} \ln\left( 1 + \frac{1}{s} \right) \right|_0^\infty - \frac{\beta}{\sigma} \int_0^\infty ds \, s^{\frac{\sigma}{\beta}} \left( -\frac{1}{s(s+1)} \right) \right] \nonumber \\ 
 = & \; - \frac{e^{\mu\sigma}}{2\pi\sigma\beta} \int_0^\infty ds \frac{s^{\frac{\sigma}{\beta}-1}}{s+1} = -\frac{e^{\mu\sigma}}{2\sigma\beta \sin\left( \pi\sigma / \beta \right)} \label{eq:int1}
\end{align}
where in the last line, we have used the integral
\beq 
\int_0^\infty ds \frac{s^{\alpha-1}}{1+s} = \frac{\pi}{\sin(\pi\alpha)}, \quad 0 < \alpha < 1. 
\eeq 
Thus, from \eq{eq:int1} and \eq{eq:int2}, we get 
\beq
I(\sigma) = -\frac{1}{2\pi\sigma^2} \left[ \frac{\frac{\pi\sigma}{\beta}}{\sin\left( \frac{\pi\sigma}{\beta} \right)} e^{\mu\sigma} - (1 + \mu\sigma) \right],   \label{eq:g_int}
\eeq 
We also note that
\beq 
\frac{1}{2\pi\sigma^2} \frac{\frac{\pi\sigma}{\beta}}{\sin\left( \frac{\pi\sigma}{\beta} \right)} e^{\mu\sigma} = \frac{1}{2\pi\sigma^2} + \frac{\mu}{2\pi\sigma} + \left( \frac{\mu^2}{4\pi} + \frac{\pi}{12 \beta^2} \right)\sigma + O(\sigma^2). 
\eeq 
Thus, the integral over the $\Theta(-\ve)$ term precisely subtracts off the singularities of the divergent integral over $g(\ve)$.

\section{Symplectic forms in noninertial frames}     \label{app:frame}
Consider the generalized Liouville 1-form for the dynamics of a classical particle on $\real^{n,1}$ with an isotropic momentum-dependent Hamiltonian:
\beq 
\eta_H = p_i dx^i - \hlt(|\vp|) dt 
\eeq 
We seek a Hamiltonian formulation of this system as seen from a noninertial frame of reference. We switch frames by a time-dependent change of coordinate $x^i = O^i_{\, j}(w^j +  \xi_j)$, where $w(t)$ corresponds to a Galilean boost and $O(t) \in \SO(n)$ to a time-dependent rotation, so that $\xi_i$ is the position coordinate in the noninertial frame. 

The derivation of a suitable symplectic form describing the dynamics in the noninertial frame then involves a choice of the definition of ``momentum''. The most straightforward choice is the \emph{canonical} momentum, defined as $\pi_i = \pi_j (O^{-1})^{\, j}_i$, which preserves the canonical ($p_idx^i$) form of $\eta_H$. We also define the velocity of the frame as $v^i = \partial_t w^i$ and its vorticity as $ (O^{-1} \partial_t O)_{ij}  = -\frac{1}{2} \omega_{ij}$, both of which may depend on time. The vorticity satisfies $\omega_{ij} = -\omega_{ji}$, which simply follows from the orthogonality of $O$.

The Liouville form becomes 
\beq
\eta_H =  \pi_i d\xi^i  - \left[ \hlt - \pi_i v^i + \frac{1}{2} \omega^{ij} \pi_i \xi_j  \right] dt \equiv \pi_i d\xi^i - \hlt' dt, 
\eeq
where assuming a slowly accelerating and rotating frame, we have only retained the terms linear in $\omega$ and $v$. Thus, for the canonical momentum, the change of frame keeps the symplectic structure invariant, while changing the Hamiltonian. In other words, $\bxi$ and $\bpi$ are canonically conjugate.

An alternative choice of momentum is the \emph{kinetic} momentum, which intends to keep the equation of motion for $\dot \xi^i$ invariant. To wit, consider the symplectic form in rotating coordinates 
\beq 
\sform_H \equiv d\rho_H = d\pi_i \wedge d\xi^i  - \left[ \frac{\partial \hlt}{\partial \pi_i} d\pi_i - v^i d\pi_i + \frac{1}{2} \omega^{ij} \left( \pi_i d\xi_j + \xi_j d\pi_i \right) \right] \wedge dt.
\eeq 
The equations of motion become 
\beq 
 \dot \xi_i = \frac{\partial \hlt}{\partial \pi_i} - v^i + \frac{1}{2} \omega^{ij} \xi_j, \quad  \dot \pi^i = - \frac{1}{2}  \omega^{ij} \pi_j. 
\eeq 
Then, one seeks the kinetic momentum $\psi^i$, in terms of which the equation of motion for $\xi$ becomes $\dot \xi^i = \partial \hlt / \partial \psi_i$. We elucidate this by examples in the following. 

\ \\ \noindent {\bf Massive case:}
Consider a massive classical particle, so that 
\[
\hlt = \frac{|\vp|^2}{2m} = \frac{|\bpi|^2}{2m} \implies  \frac{\partial \hlt}{\partial \psi_i} = \frac{\pi_i}{m}.
\]
Then, the kinetic momentum is defined by setting  
\beq 
 \dot \xi_i = \frac{\psi_i}{m} \implies \pi^i = \psi_i + m v_i - \frac{1}{2} m \, \omega^{ij} \xi^j,
\eeq 
so that 
\begin{align*}
 \hlt' 
 = & \; \frac{1}{2 m} \left| \psi_i + m v_i - \frac{1}{2} m \, \omega^{ij} \xi^j \right|^2 - \left( \psi_i + m v_i - \frac{1}{2} m \, \omega^{ij} \xi^j \right) \left( v^i - \frac{1}{2} \omega^{ik} \right) \\ 
 = & \; \frac{\psi^2}{2 m^2} + \psi_i v^i - \frac{1}{2} \omega^{ij} \psi_i \xi_j - \psi_i v^i + \frac{1}{2} \omega^{ij} \psi_i \xi_j + \text{second order terms} \\
 = & \; \frac{\psi^2}{2 m^2} + \text{second order terms}.
\end{align*}
It is precisely this cancellation that we seek in defining the kinetic momentum. Thus to linear order in $v$ and $\omega$, defining $a^i = \partial_t v^i$ and $\alpha_{ij} = \partial_t \omega_{ij}$, the symplectic form becomes 
\beq
 \sform_H = d\psi_i \wedge d\xi^i + \frac{1}{2} m \, \omega_{ij} d\xi^i \wedge d\xi^j + m \left(a_i + \alpha_{ij} \xi^j \right) dt \wedge d\xi^i - d \left( \frac{|\bps|^2}{2m} \right) \wedge dt.
\eeq
We combine the inertial terms as
\beq
\Omega = \frac{1}{2} \Omega_{\mu\nu} dx^\mu dx^\nu = \frac{1}{2} \omega_{ij} d\xi^i \wedge d\xi^j + \left(a_i + \alpha_{ij} \xi^j \right) dx^0 \wedge d\xi^i, 
\eeq
the symplectic form simply becomes 
\beq 
\sform_H = d\psi_i \wedge d\xi^i + m \, \Omega - d\hlt \wedge dt.
\eeq 
Here, $m \omega$ corresponds to the Coriolis force, $m a$ to the inertial force and $m \alpha_{ij} \xi^j dt dx^i$ to the tangential acceleration due to a variable angular velocity. This does not capture the centrifugal force, as we have ignored the terms at $O(\omega^2)$.

\ \\ \noindent {\bf Massless case:}
For massless particles, $\hlt = \vel|\vp| = \vel|\bpi|; \; \vel = \pm 1$, so that the equations of motion become 
\[ 
 \dot \xi_i = \vel \hat \pi^i - v^i - \frac{1}{2} \omega^{ij} \xi_j, \quad  \dot \pi^i =  \frac{1}{2}  \omega^{ij} \pi_j. 
\]
Taking a cue from the massive case, consider a definition of kinetic momentum as 
\beq 
\pi^i = \psi_i + c |\bps| \left( v_i - \frac{1}{2} \omega^{ij} \xi^j \right); \quad c^2 = 1.
\eeq 
Expanding 
\[
 \hlt' = \vel |\bpi|= \vel |\bps| \left[ 1 + 2 \vel \hat\psi^i \left( v_i - \frac{1}{2} \omega^{ij} \xi^j \right) \right]^{1/2} = \vel |\bps| +  \psi^i \left( v_i - \frac{1}{2} \omega^{ij} \xi^j \right) + O(\Omega^2),
\]
we again get a cancellation in $\hlt'$ at linear order, so that $\hlt' = \vel |\bps|$. Thus, the symplectic form becomes 
\beq 
\sform_H = d\psi_i \wedge d\xi^i + \vel |\bps| \, \Omega - \vel \, d|\bps| \wedge dt.
\eeq 
Note that in considering the Galilean boosts (instead of Lorentz boosts), we are ignoring the effect of time dilation, including which will lead to corrections at the next order in $\Omega$.

\section{No-drag frame}   \label{app:no_drag}
In this appendix, we show that the frame with respect to which our Weyl fluid is in equilibrium satisfies the ``no-drag'' condition described by Stephanov and Yee\cite{misha_no_drag}. In $3+1$ dimensions, the grand potential current becomes\footnote{Note that there are missing factors of $2\pi$ in the expansions of $\bg_{anom}$ in Ref. \onlinecite{loga_weyl_gas}, eq. A.12 - A.15, as the expansion should always have the combination $\frac{qB}{2\pi}$ instead of $qB$.} 
\beq 
 \bg = - \left( \frac{\mu^3}{24\pi^2} + \frac{\mu T^2}{24}  \right) u \wedge \omega - \left( \frac{\mu^2}{8\pi^2} + \frac{T^2}{24}  \right) u \wedge B, 
\eeq 
so that 
\begin{align}
 \bcc = & \; -\frac{\partial \bg}{\partial \mu} =  \left( \frac{\mu^2}{4\pi^2} + \frac{T^2}{12}  \right) u \wedge \omega_{A} + \frac{\mu}{4\pi^2}  u \wedge B \nonumber \\  
 \bec = & \; -\frac{\partial \bg}{\partial T} = \frac{\mu T}{6}  u \wedge \omega_{A} + \frac{T}{12}  u \wedge B \nonumber \\ 
 \bq = & \; \bg + \mu \bcc + T \bec =  \left( \frac{\mu^3}{6\pi^2} + \frac{\mu T^2}{6}  \right) u \wedge \omega_{A} + \left( \frac{\mu^2}{8\pi^2} + \frac{T^2}{24}  \right) u \wedge B, 
\end{align}
where $\omega_{A} = \frac{1}{2} \omega$. Thus, we can identify the coefficients 
\[
\begin{array}{lll}
 \xi_{J, \omega} = \frac{1}{12} T^2 + \frac{1}{4\pi^2}\mu^2, \quad  & 
 \xi_{S, \omega} = \frac{1}{6} \mu T, \quad   & 
 \xi_{T, \omega} = \frac{1}{6}\mu T^2 + \frac{1}{6\pi^2}\mu^3,  \\
 \xi_{J, B} =  \frac{1}{4\pi^2}\mu, & 
 \xi_{S, B} =  \frac{1}{12}T, & 
 \xi_{T, B} = \frac{1}{24} T^2 + \frac{1}{8\pi^2} \mu^2.
\end{array}
\]
Comparing with the expressions obtained in Ref \cite{misha_no_drag} 
\[
\begin{array}{lll}
 \xi_{J, \omega} = X_B T^2 + C \mu^2,  \quad & 
 \xi_{S, \omega} = X_\omega T^2 + 2 X_B \mu T, \quad & 
 \xi_{T, \omega} = \frac{2}{3} \left( X_\omega T^3 + 3 X_B \mu T^2 + C \mu^3 \right) \\
 \xi_{J, B} = C \mu, &
 \xi_{S, B} =  X_B T, & 
 \xi_{q, B} = \frac{1}{2} \left( X_B T^2 + C \mu^2 \right),
\end{array}
\]
we can readily identify 
\beq 
C = \frac{1}{4\pi^2}, \quad X_B = \frac{1}{12}, \quad X_\omega = 0. 
\eeq 
Thus, our transport coefficients are indeed consistent with the no-drag frame.

\bibliography{anom_fluid}

\end{document}